\documentclass[
    ,final            
    ,numberedheadings 
    ,cmfonts          
  ]
  {aipproc}

\usepackage{graphicx}

\layoutstyle{6x9}

\begin{document}

\title{
Probing the partonic structure of exotic particles in hard electroproduction
}

\classification{12.38.Bx 12.39.Mk}
\keywords      {Hybrid meson electroproduction}

\author{I.V.~Anikin}{
  address={LPT, Universit\'e Paris-Sud$\,$\footnote{Unit\'e mixte 8627 du CNRS}
, F-91405 Orsay, France},
altaddress={CPhT, \'Ecole Polytechnique$\,$\footnote{Unit\'e mixte 7644 du CNRS}, F-91128 Palaiseau, France}
}

\author{B.~Pire}{
  address={CPhT, \'Ecole Polytechnique$\,$\footnote{Unit\'e mixte 7644 du CNRS}, F-91128 Palaiseau, France}
}

\author{L.~Szymanowski\footnote{E-mail: lechszym@fuw.edu.pl} \   }{
  address={Universit\'e de  Li\`ege, Sart Tilman,B-4000 LIEGE 1, Belgium},
altaddress={SINS, Warsaw, Poland}
}

\author{O.V.~Teryaev}{
  address={BLTP,  Joint Institute for Nuclear Research, Dubna, Russia}
}

\author{S.~Wallon}{
  address={LPT, Universit\'e Paris-Sud$\,$\footnote{Unit\'e mixte 8627 du CNRS}
, F-91405 Orsay, France}
}

\begin{abstract}

We argue that the electroproduction of exotic particles
is a useful tool for study of their  partonic structure.  In
the case of  hybrid mesons, the magnitude of their   cross sections shows that they
are accessible for measurements
in existing electroproduction experiments.

\end{abstract}

\maketitle


\section{Introduction}

Searching for exotic particles whose structure cannot be explained in the framework of
 the constituent quark model is at  present a very lively subject of studies.
One of their main streams concentrates on finding a firm evidence for the existence 
(or absence) of pentaquarks, with the lowest Fock state containing 
$uudd\bar s$ quarks. 
This is particularly important in view of some contradictory results 
of experimental searches, reviewed at this conference by I.~Strakovsky \cite{Strakovsky}.
Because of the apparently very small decay width of pentaquarks ($\approx 1\,$MeV) 
their theoretical description represents a big challenge, as discussed 
 by M.~Polyakov \cite{Polyakov}. 
Another family of exotic particles which are at present a subject of intense 
studies are
glueballs and hybrid mesons, for their review see, e.g. \cite{hybexp}.
 The theoretical description of 
these particles requires to include in the lowest Fock state the 
gluonic degrees of freedom. All these facts stress the importance of studies of the 
 structure of exotic particles within Quantum Chromodynamics (QCD) for understanding
the  mechanism of quark and gluon confinement.

The hard exclusive electroproduction of particles permits to probe the structure of 
particles at the parton level. The scattering amplitude factorizes in this case into  
perturbatively calculable coefficient function, the non-perturbative 
distribution amplitude (DA) of a produced particle and the non-perturbative generalized 
parton distribution (GPD) describing the transition probability amplitude of the target.
The DA and GPD encode full information about parton distribution inside
particles participating in a collission, for a review see, e.g. \cite{Diehl}.

Recently it was emphasized \cite{penta} that 
 hard electroproduction can also be used to reveal  the partonic structure 
of exotic particles. In particular, for the processes involving pentaquarks
\begin{equation}
\label{procpenta}
\gamma^* p \to \bar K^0 \Theta^+ \;,\;\;\;\;\;\;\;\gamma^* n \to \bar K^- \Theta^+\;,
\end{equation}
as well as for their analogs with  vector mesons $K^*$,
 the necessary formalism based on the
GPD of $p(n) \to \Theta^+$ was developed which has permitted next to evaluate some contributions to the
 cross section 
in the Born approximation \cite{penta}. Unfortunately,  realistic cross section 
estimates based on these expressions are not possible at the moment due to our 
ignorance of these  $p(n) \to \Theta^+$ GPDs. 
 
Another example of processes in which the partonic structure of exotic particles can be probed 
is supplied by
the hard electroproduction of  $J^{PC}=1^{-+}$ hybrid meson $\pi_1(1400)$ 
\begin{equation}
\label{proces}  
\gamma^* p \to \pi_1 p\;.
\end{equation}
The process (\ref{proces}) is the  subject of  recent
studies in Refs. \cite{hybRC} and \cite{hybPRD}. Below we present 
in some details our results.

\section{The puzzle with DA of  $J^{PC}=1^{-+}$ hybrid meson}

In  Ref. \cite{JJZ} Jaffe {\it et al.} analyse the  particle spectrum (including its exotic sector)  
by  construction 
 of lowest-dimensional, gauge invariant, colorless {\it local}  
operators. As a result of this study
  a hybrid meson with exotic quantum numbers $J^{PC}=1^{-+}$
is described by a {\it local} composite operator of 
dimension 5 built from quark  
and gluonic fields, e.g. 
$\bar \psi \gamma^\mu G_{\mu\,\nu}\,\psi $, where 
$G^{\mu\,\nu}$ is the field strength tensor of a gluon.
Such local operator has a twist equal to 4. 
Naively one could think therefore that this result implies that the leading twist-2
DA of a hybrid meson vanishes. Consequently, the scattering amplitude for hard
hybrid meson electroproduction would be  
suppressed (at large photon virtuality $Q^2$) in comparison with the 
amplitude for production in the same process of non-exotic 
$\rho-$meson ($J^{PC}=1^{--}$).
The analysis of Ref.~\cite{hybRC} shows that this naive conclusion is not correct
for longitudinally polarized hybrid meson for
which the leading twist-2 DA  is not zero. Thus, the 
electroproduction of such hybrid meson doesn't need to be strongly suppresed 
in comparison 
with a $\rho-$meson production. 

The analysis of  Ref.~\cite{hybRC} 
consists in exploiting the fact  that DAs of particles are defined by a {\it non local} composite
 operators. In the case of a longitudinally polarised hybrid meson $H(p,0)$ 
of momentum $p$ 
the DA has the form
\begin{equation}
\langle H(p,0)| \bar \psi(-z/2)\gamma_\mu [-z/2;z/2]
\psi(z/2) | 0 \rangle =
i f_H M_H e^{(0)}_{L\,\mu}
\int\limits_0^1 dy e^{i(\bar y - y)p\cdot z/2} \phi^{H}_L(y)\;,
\label{hybDA}
\end{equation}
where $[-z/2;z/2]$ on the l.h.s. 
denotes the path-ordered gluonic exponential along the 
straight line connecting
the initial and final points $(-z/2,z/2)$ which provides the gauge invariance
for bilocal operator and
equals unity in a light-like (axial) gauge.
In Eq.~(\ref{hybDA}) $f_H$, $M_H$, $e^{(0)}_\mu$ denote the hybrid meson 
coupling constant, its mass and the longitudinal polarisation vector, respectively.
Because of the positive charge parity of the hybrid meson, 
 its DA, $\phi^H$, 
is an antisymmetric function: $\phi^{H}(y)=-\phi^{H}(1-y)$. This last property implies
in particular that 
\[
\int\limits_{0}^{1}dy \; \phi^{H}(y)=0\;,
\]
which means that the first term of the Taylor expansion in $z$ of l.h.s of (\ref{hybDA})
vanishes and generally only terms with odd powers of $z$ contribute to this expansion 
\begin{eqnarray}
\label{locdec}
&&\langle H(p,\lambda)| \bar\psi(-z/2) \gamma_{\mu}[-z/2;z/2] \psi(z/2)| 0\rangle=
\\
&&\sum_{n\, odd}\frac{1}{n!}z_{\mu_1}..z_{\mu_n} \langle H(p,\lambda)|
\bar\psi(0) \gamma_{\mu}
\stackrel{\leftrightarrow}{D}_{\mu_1}..\stackrel{\leftrightarrow}{D}_{\mu_n}
\psi(0)| 0\rangle ,
\nonumber
\end{eqnarray}
in which  $D_{\mu}$ is the usual covariant derivative and
$\stackrel{\leftrightarrow} {D_{\mu}}=\frac{1}{2}(
\stackrel{\rightarrow}{D_{\mu}}-
\stackrel{\leftarrow}{D_{\mu}}$).
The first non-vanishing term of the expansion (\ref{locdec}) corresponds to $n=1$ and its twist-2
contribution involves operator
\begin{equation}
{\cal R}_{\mu\nu}=\mbox{\cal S}_{(\mu \nu)}
\bar\psi(0)\gamma_{\mu}
\stackrel{\leftrightarrow}{D}_{\nu}\psi(0),
\end{equation}
where ${\cal S}_{(\mu\nu)}$ denotes the  symmetrization
operator (${\cal S}_{(\mu \nu)}T_{\mu \nu}=1/2(T_{\mu \nu}+T_{\nu
\mu})$).
Let us note that
${\cal R}_{\mu \nu}$ is
proportional to the quark energy-momentum tensor,
 ${\cal R}_{\mu\nu} = - i \Theta_{\mu\nu}$.
Its matrix element of interest is
\begin{equation}
\label{emee}
\langle H(p,\lambda) | {\cal R}_{\mu\nu} |0\rangle=
 \frac{1}{2}\,f_H M_H \mbox{\cal S}_{(\mu \nu)}
e_{\mu}^{(\lambda)} p_{\nu} \int\limits_{0}^{1}dy (1-2y) \phi^{H}(y).
\end{equation}
Examining symmetry properties of the operator ${\cal R}_{\mu \nu}$ and the matrix element 
(\ref{emee})
reveals indeed that the $C$ and $P$ parities of  ${\cal R}_{\mu \nu}$
are $(-1)$, equal to those of the hybrid meson. This proves that $f_H$ is non zero and allows to 
determine its value through non perturbative methods, such as, e.g. the QCD sum rules method
\cite{Bal}: \\
a) using the equations of motion one derives that
$\partial^\mu \Theta_{\mu \nu} = g \bar \psi \gamma^\mu G_{\mu \nu} \psi\;, $ \\
b) the value of the coupling constant $f_H$ is determined by
 the correlator of two  
$\bar \psi \gamma^\mu G_{\mu \nu} \psi$ operators. \\
This results in the estimate
$ f_{H } \approx 50 \,{\rm MeV}\;.$

The description of the DA of hybrid meson is complete by fixing the form of $\phi^H(y)$. This DA satisfies
usual non-singlet evolution equations and,
forgetting the slowly varying logarithmic scaling violation
factor,
 we assume in our estimates that it is given by 
 the  asymptotic expression \cite{Chase}: 
 $\phi^H(y)_{as}=30y(1-y)(1-2y)$.

\begin{figure}
  \includegraphics[height=.3\textheight]{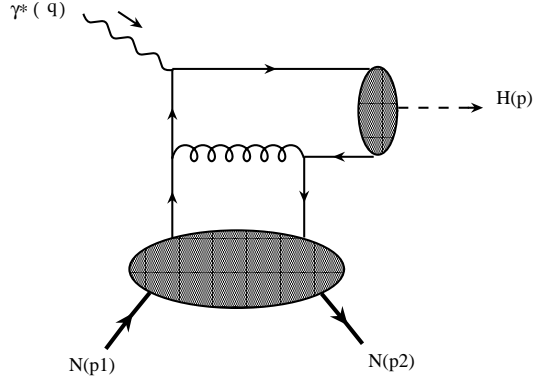}
  \caption{Typical diagram describing the electroproduction of a meson at lowest order. The grey 
blobs are the nonperturbative  meson distribution amplitude and  nucleon generalized parton 
distribution.}
\end{figure}

\section{The hybrid meson production}

\subsection{The scattering amplitude}

Knowing  the DA of the hybrid meson (\ref{hybDA}) one can determine the scattering amplitude
at the leading order in  strong coupling constant $\alpha_s$. The calculations proceed
in a full analogy as in the $\rho$-meson case.
The electroproduction process (\ref{proces}) occurs in the scaling regime where the virtuality
of the photon $Q^2=-q^2$, see Fig.~1, is large and scales with the energy of the process, 
and the momentum transfer $t$ is small, $-t << Q^2$.
In such kinematics the conditions for  the QCD factorization are fulfilled and the 
scattering amplitude at leading twist is at given factorisation scale $\mu$ expressed as a convolution
\begin{equation}
\label{amplit}
{\cal A}=\int\limits_0^1du\;\int\limits_0^1dx\;\phi(u,\mu^2)H(x,u,Q^2,\mu^2,\mu_R^2)F(x,\mu^2)
\equiv \phi \otimes H \otimes F
\end{equation}   
of the DA $\phi(u,\mu^2)$ of meson, the GPD of the target nucleon  $F(x,\mu^2)$ and the perturbatively 
calculable coefficient function  $H(x,u,Q^2,\mu^2,\mu_R^2)$. The hard part of the scattering amplitude
depends also on the renormalisation scale $\mu_R$. The expression for $H(x,u,Q^2,\mu^2,\mu_R^2)$
is obtained by adding the contributions of 4 diagrams, one of which is shown in Fig.~1. Taking into account 
the contributions of all diagrams one arrives to the scattering amplitude of the form
\begin{equation}
\label{qdsim2}
{\cal A}_{\gamma^*_L p\to H^0_L p}=  \frac{e\pi\alpha_s
f_{H}\,C_F}{\sqrt{2}\;N_c\,Q}
 \biggl[ e_u {\cal H}_{uu}^- -e_d {\cal H}_{dd}^-\biggr] {\cal V}^{H\,-},
\end{equation}
where
\begin{eqnarray}
\label{softin}
{\cal H}_{ff}^\pm=
\int\limits_{-1}^{1}dx &&\biggl[
\overline{U}(p_2) \hat{n} U(p_1) H_{ff}(x,\xi) +
\overline{U}(p_2)\frac{i\sigma_{\mu\alpha}
n^{\mu}\Delta^{\alpha}}{2M}U(p_1)
E_{ff}(x,\xi) \biggr]
\nonumber \\
&&
\times \, \biggl[
\frac{1}{x+\xi-i\epsilon} \pm \frac{1}{x-\xi+i\epsilon}\biggr],
\end{eqnarray}
and
\begin{eqnarray}
{\cal V}^{M,\pm} =
\int\limits_{0}^{1} dy \phi^{H}(y)\biggl[
\frac{1}{y}-\frac{1}{1-y}
\biggr].
\nonumber
\end{eqnarray}
The functions $H$ and $E$ are standard leading twist GPD's having well known properties. 
Eq.~(\ref{softin}) also show definitions of ${\cal H}_{ff}^+$ and ${\cal V}^{M,\,+}$  necessary for 
comparison with the production of $\rho$-meson.

\subsection{Scale fixing and numerical predictions.}

The QCD factorisation of the scattering amplitude as given by Eq.~(\ref{amplit}) introduces 
 dependence of
the coefficient function $H$, the soft DA $\phi$ and the target GPD $F$ on the 
factorization scale $\mu$ and on the renormalisation scale $\mu_R$, which in principle 
should be treated as two independent parameters. Since in the coefficient function $H$ 
only first terms of its perturbative expansion are at best known, the 
dependence of the amplitude ${\cal A}$
on $\mu$ and $\mu_R$ can be large and leads to significant theoretical uncertainties of 
 results. In order to minimalize this uncertainty a scale fixing procedure has to be
invoked. 
 
Inspired by results obtained in calculations of the pion form-factor we adapted the convention that
 both scales are equal, $\mu=\mu_R$ \cite{hybPRD}. Fixing of the scale $\mu$ 
is then done by 
applying a modified version of the Brodsky-Lepage-McKenzie (BLM) procedure: the scale $\mu$ is 
chosen in such a way which leads to a vanishing of large terms proportional to the $\beta$-function    
(which governs the $\mu$-behaviour of the strong coupling constant $\alpha_s(\mu^2)$) in the square of 
the scattering 
amplitude known with at least  the next-to-leading order accuracy, see \cite{BLM} for details.
In more physical terms the BLM procedure consists in absorbing numerically large terms 
originating from the renormalisation into redefinition of the argument of the strong coupling constant
$\alpha_S(\mu^2)$.    

Results of the numerical analysis of hybrid meson 
electroproduction are shown in Fig.~2. The solid line represents the cross-section
 of the non-exotic  $\rho$-meson electroproduction (quark exchange contribution only) 
obtained with the BLM scale fixing. The dashed line
describes the cross section for  production of the hybrid meson $\pi_1(1400)$. 
The comparison of these two
curves leads to the conclusion that the production process of hybrid 
meson has a sizeable cross section, so that it  should be already now accessible to
measurements.
Fig.~2 also contains  the comparison of our results with the predictions of Ref.~\cite{Vand99}
on the $\rho$-meson production which takes into account the intrinsic transverse momenta of partons
without invoking the BLM procedure.

\begin{figure}[h]
$$\rotatebox{270}{\includegraphics[width=8cm]{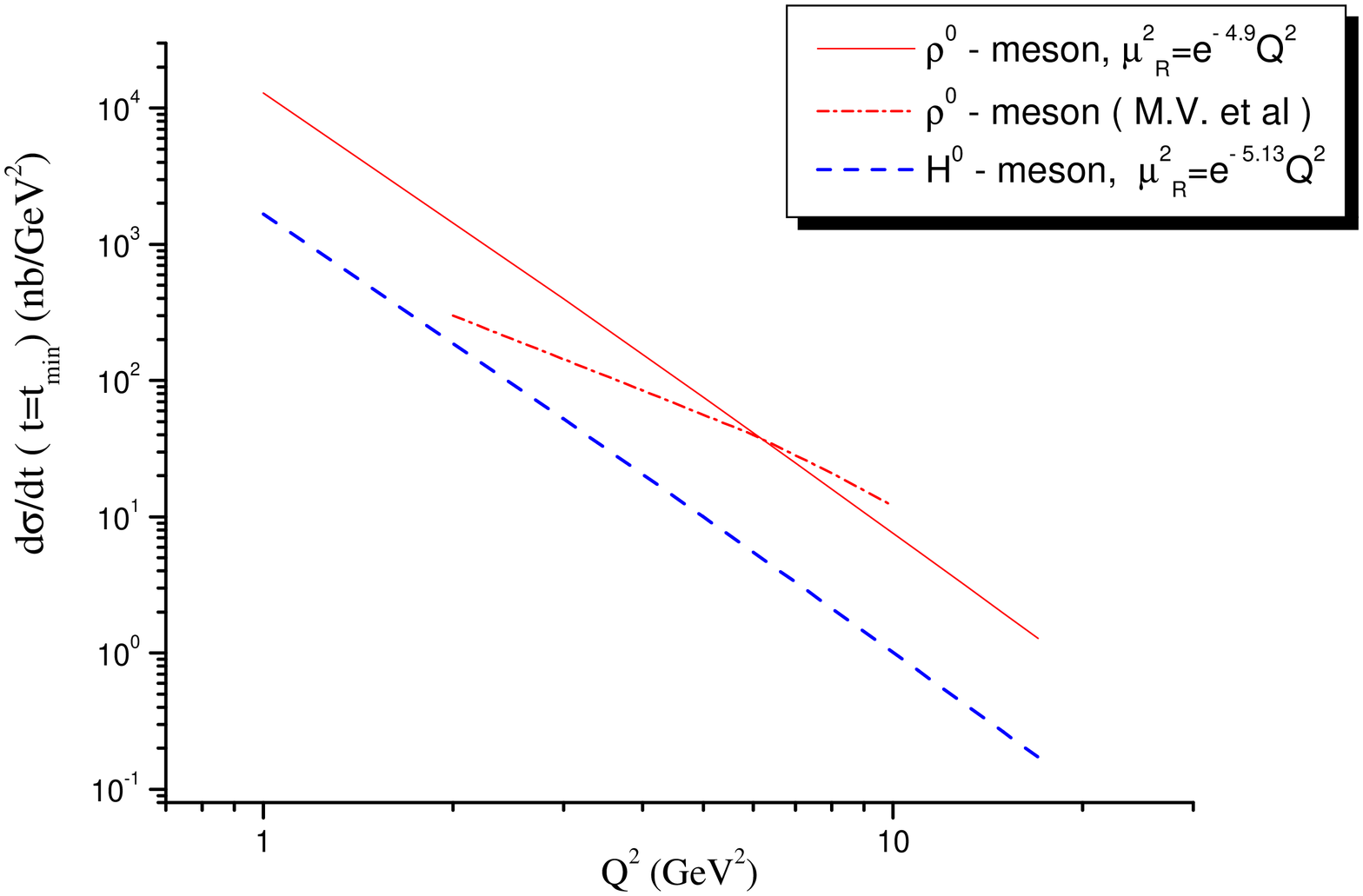}}$$
\caption{
Differential cross-section for exotic hybrid meson
electroproduction (dashed line) with
$\mu_R^2 =e^{-5.13} Q^2$ compared with
the quark contribution to $\rho^0$
electroproduction (solid line) with $\mu_R^2 =e^{-4.9} Q^2$, as a function
of $Q^2,$ for $x_B\approx 0.33$. The dash-dotted line is the result of Vanderhaegen et al 
\cite{Vand99} for $\rho$ electroproduction.
}
\label{conaiv}
\end{figure}


\section{Hybrid meson probed through the electroproduction of $\pi\eta$ pairs}

\begin{figure}[h]
\label{diag_hyb}
  \includegraphics[height=.3\textheight]{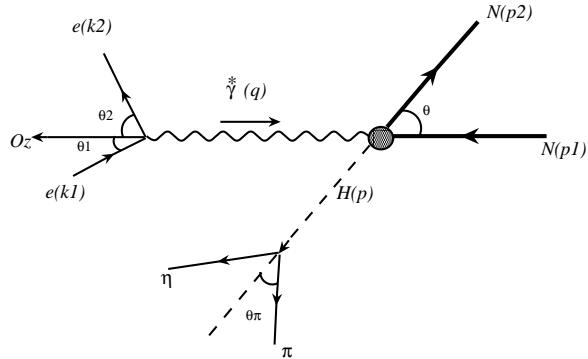}
 \caption{The kinematics of the electroproduction of $\pi\eta$ pair}
\end{figure}
\begin{figure}[h]
\label{hyb_f2}
  \includegraphics[height=.3\textheight]{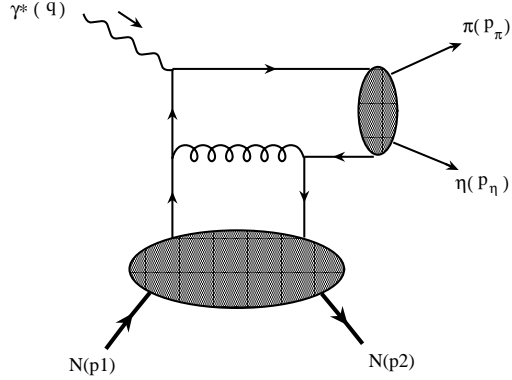}
  \caption{Typical diagram describing the electroproduction of $\pi\eta$ pair. The grey 
blobs are non-perturbative the $\pi\eta$-pair GDA and the nucleon GPD.}
\end{figure}

In the case where there is no recoil detector which allows to identify the
hybrid meson production events through a missing mass reconstruction, one has to base
an identification process through the possible decay products of the hybrid meson $H^0$. Since the
particle $\pi_{1} (1400)$  has a dominant $\pi\eta $ decay mode one can use the electroproduction 
process (see Fig.~(\ref{diag_hyb})),
\begin{eqnarray}
\label{eN}
e(k_1)+N(p_1)\to e(k_2)+\pi^0(p_\pi)+\eta(p_\eta) + N(p_2)
\end{eqnarray}
or
$
\gamma^*(q)+N(p_1)\to \pi^0(p_\pi)+\eta(p_\eta) + N(p_2)
$, (see Fig.~(\ref{hyb_f2})),
to probe the  hybrid meson properties. 
The computation of the process in Fig.~(\ref{hyb_f2})
requires a knowledge of  the generalized distribution
amplitude (GDA) for the $\pi\eta$ pair \cite{DGPT}:
\begin{eqnarray}
\label{hme2}
&&\langle \pi^0(p_\pi)\eta(p_\eta) |
\bar\psi_{f_2}(-z/2)\gamma^{\mu}[-z/2;z/2] \tau^3_{f_{2}f_{1}}
\psi_{f_1}(-z/2)|0\rangle=
\nonumber\\
&&p^{\mu}_{\pi\eta}\int\limits_{0}^{1}dy e^{i(\bar y-y)p_{\pi\eta}\cdot z/2}
\Phi^{(\pi\eta)}(y,\zeta, m_{\pi\eta}^2),
\end{eqnarray}
where  the total momentum of $\pi\eta$ pair is $p_{\pi\eta}=p_{\pi}+p_{\eta}$
while $m^2_{\pi\eta}=p^2_{\pi\eta}$. The $\pi\eta$ distribution amplitude 
$\Phi^{(\pi\eta)}$ describes non resonant
as well as resonant contributions. It does not possess any symmetry properties
concerning the $\tilde \zeta$-parameter
\begin{equation}
\label{zeta}
\tilde\zeta=\frac{p_\pi^+}{(p_\pi+p_\eta)^+}-
\frac{m^2_\pi-m^2_\eta}{2m^2_{\pi\eta}},\;\;\;\;\;\;\;
1-\tilde\zeta=\frac{p_\eta^+}{(p_\pi+p_\eta)^+}+
\frac{m^2_\pi-m^2_\eta}{2m^2_{\pi\eta}},
\end{equation}
which describes roughly the  fraction of total `+' momentum of the pair carried by the $\pi$-meson 
(in case of particles with equal masees $\tilde \zeta=p_{\pi}^+ / p^+$) and which is related to 
the angle $\theta_{cm}¥$,
defined as the polar angle of the $\pi$ meson in the center of mass frame of the meson pair:
\begin{eqnarray}
\label{2zeta}
2\tilde\zeta-1=\beta\cos\theta_{cm}\,,\;\;\;\;\;\;\;\;\;\;\;
\beta=\frac{2|{\bf p}|}{m_{\pi\eta}},
\end{eqnarray} 
In the reaction under  study, the $\pi\eta$ state may have
total momentum, parity and charge-conjugation in the following
sequence $J^{PC}=0^{++},\, 1^{-+},\, 2^{++}, \, ...$,
 that corresponds to the following values of the $\pi\eta$ orbital angular momentum $L$:
$L=0,\, 1, \, 2,\, ...$,
respectively. Thus a resonance with a $\pi\eta$ decay mode for odd
orbital angular momentum $L$ should be considered as an exotic meson. 
The mass region around $1400$ ${\rm MeV}$ is dominated by the strong $a_2(1329)\,(2^{++})$
resonance, it is therefore natural to look for the
interference of the amplitudes of hybrid and $a_{2}$ production, resulting in the angular asymmetry in
$\pi\eta$ production.

Asymmetries are often a good way to get a measurable signal for a small
amplitude by means of its interference with a larger one. 
In the asymmetry a small amplitudes enters 
linearly rather than quadratically as in the cross section which increases chances for sizeable effects. 
 In our case,
since the hybrid production amplitude may be rather small with respect to
a continuous background, we  use the supposedly
large amplitude for $a_{2}$ electroproduction as a magnifying lens to unravel
the presence of the exotic hybrid meson. Since these two amplitudes describe
different orbital angular momentum of the $\pi$ and $\eta$ mesons, the asymmetry
which is sensitive to their interference is an angular asymmetry  defined by
\begin{eqnarray}
\label{anas}
 A(Q^2, y_l,\hat t, m_{\pi\eta})=
 \frac{\int \cos\theta_{cm} \,
d\sigma^{\pi^0\eta}(Q^2, y_l,\hat t, m_{\pi\eta}, \cos\theta_{cm} )}{
\int d\sigma^{\pi^0\eta}
(Q^2, y_l,\hat t, m_{\pi\eta}, \cos\theta_{cm} )}
\end{eqnarray}
as a weighted integral over polar angle $\theta_{cm}$ of the relative momentum of
$\pi$ and $\eta$ mesons. The variable $y_l$ is the longitudinal fraction of the electron momentum $k_1$
carried by the virtual photon.  

\begin{figure}[h]
$$\rotatebox{270}{\includegraphics[width=8cm]{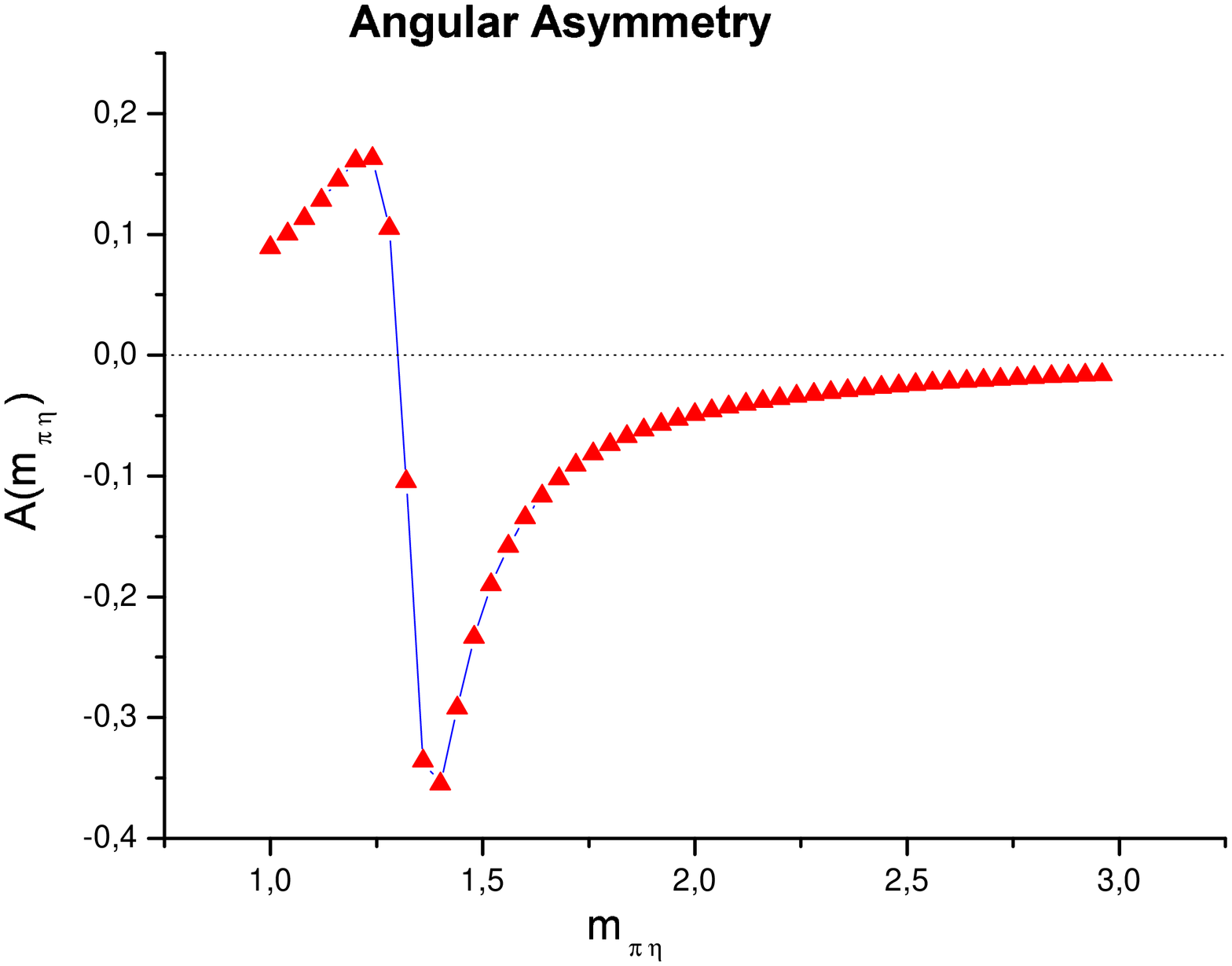}}$$
\caption{The angular asymmetry as a function of $m_{\pi\eta}$.}
\label{angular}
\end{figure}

Our estimation of the asymmetry (\ref{anas}) is shown on Fig. \ref{angular} and it has a 
sizable magnitude.  The structure of this asymmetry is very natural:  
 the first positive extremum
is located at $m_{\pi\eta}$ around the mass of $a_2$ meson while the second negative extremum
corresponds to the hybrid meson mass.

\section{Summary}

We advocate to use  hard electroproduction to uncover the
 partonic structure of exotic particles. In the case of  the  exotic  $J^{PC}=1^{-+}$ hybrid meson 
$\pi(1400)$ we presented quantitative estimates of
the leading twist contribution to the electroproduction
amplitude, which lead  to sizeable effects.  The resulting order of magnitude
of cross section is  smaller than the $\rho$ electroproduction but
similar to the $\pi$ electroproduction. Thus the exotic hybrid meson effects 
 should be measurable at dedicated experiments at JLab, Hermes or Compass.


\begin{theacknowledgments}
 This work is supported by the Polish Grant 1 P03B 028 28, the Joint 
Research Activity ``Generalized Parton Distributions'' of the european 
I3 program Hadronic Physics, contract RII3-CT-506078 and the French-Polish scientific agreement 
Polonium.
The work of I.V.A. and O.V.T. is supported in part by INTAS (Project 00/587)
and RFBR (Grant 03-02-16816). I.~V.~A. thanks  NATO for a Grant. 
L.Sz. is a 
Visiting Fellow of the Fond National pour la Recherche Scientifique 
(Belgium).

\end{theacknowledgments}

\end{document}